\documentclass{article}
\usepackage{spconf}

\usepackage{cite}
\usepackage{amsmath,amssymb,amsfonts}
\usepackage{graphicx}
\usepackage{textcomp}
\usepackage{xcolor}
\newcommand{\revisiontext}[1]{{#1}}
\usepackage{booktabs}
\usepackage{etoolbox}
\makeatletter
\patchcmd{\@maketitle}{\vskip 1.5em}{\vskip 1em}{}{}
\patchcmd{\@maketitle}{\vskip 1.5em}{\vskip 1em}{}{}
\makeatother
\usepackage[hidelinks]{hyperref}
\IfFileExists{pcrr7t.tfm}{}{}

\def\BibTeX{{\rm B\kern-.05em{\sc i\kern-.025em b}\kern-.08em
    T\kern-.1667em\lower.7ex\hbox{E}\kern-.125emX}}

\begin{document}
\small
\renewcommand{\baselinestretch}{1}
\renewcommand{\normalsize}{\fontsize{9.5pt}{10.45pt}\selectfont}
\normalsize

\title{Factorized Delayed Streams Modeling for LLM-based Streaming ASR}

\name{\shortstack{Tatsunari Takagi$^{1,2}$, Kai Washizaki$^{1}$, Atsushi Kojima$^{1}$, Lianbo Liu$^{1}$, Koki Nikaido$^{1}$, Yui Sudo$^{1}$}}
\address{$^{1}$SB Intuitions \\
         $^{2}$Toyohashi University of Technology}

\maketitle

\begin{abstract}
Delayed Streams Modeling (DSM) enables LLM-based streaming automatic speech
recognition (ASR) by aligning acoustic and text streams on a common timeline.
DSM adds the padding token \texttt{<p>} and the word-start token \texttt{<w>} to
the LLM vocabulary and predicts them together with normal text tokens using the
same softmax. We first show that \texttt{<w>} can be removed while maintaining
competitive recognition performance. Based on this result, we propose Factorized
DSM (F-DSM), which separates the waiting probability for \texttt{<p>} from the
distribution over the original LLM vocabulary. This factorization removes
ASR-specific tokens from the text prediction space and allows the
large-vocabulary softmax to be skipped on waiting steps. Experiments on the
Corpus of Spontaneous Japanese and LibriSpeech show that F-DSM achieves better
recognition performance than DSM. It also greatly reduces GPU memory use while
maintaining similar training throughput, provides a small inference speed
improvement through softmax skipping, and reduces the degradation in text-only
perplexity observed with DSM.

\end{abstract}


\begin{figure*}[t]
    \centering
    \includegraphics[width=0.86\textwidth]{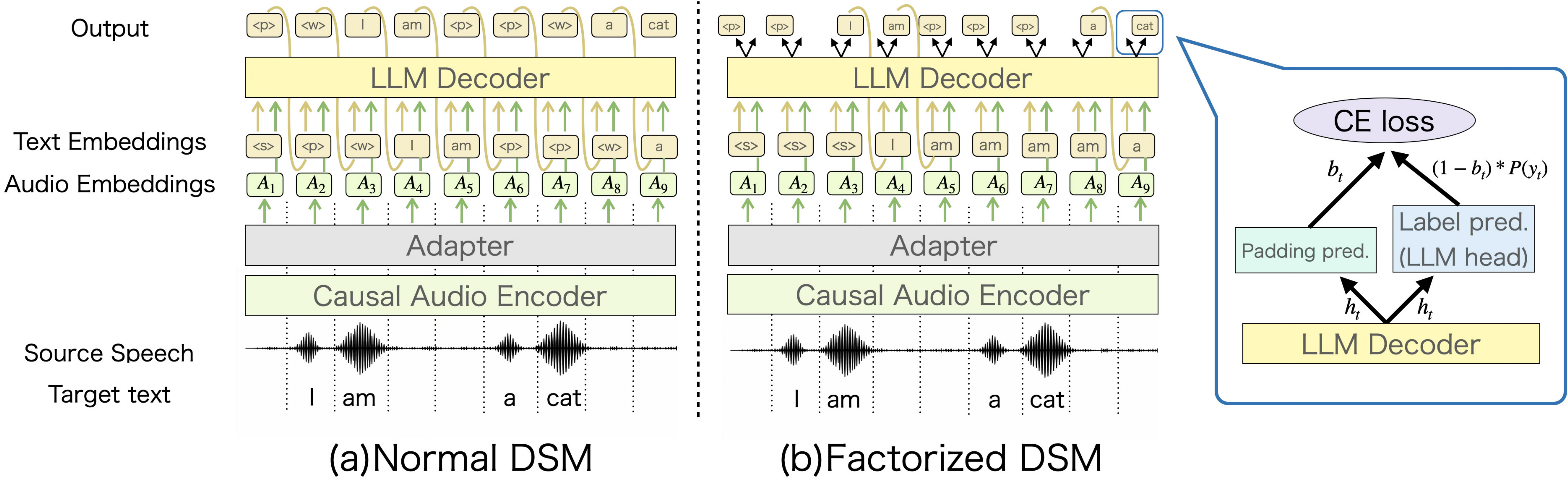}
    \caption{DSM (left) jointly predicts waiting \texttt{<p>}, word-start
    \texttt{<w>}, and text tokens; F-DSM (right) separates waiting from
    text prediction over the original LLM vocabulary.}
    \label{fig:method-overview}
\end{figure*}

\section{Introduction}

\label{sec:intro}

Streaming automatic speech recognition (ASR) processes incoming speech step by
step and produces text as soon as enough acoustic information is available
\cite{StreamingE2E,cache_aware_streaming}. At each step, the model must decide
whether to output a text token or wait for more speech. In Connectionist
Temporal Classification (CTC) \cite{CTC} and Recurrent Neural Network-Transducer
(RNN-T) \cite{RNNT}, this ``no output'' state is modeled by a blank symbol,
which plays a different role from normal text tokens.

Recently, large language models (LLMs) \cite{LLM_overview,LLM_survey} have been
widely studied for ASR because of their strong language modeling ability
\cite{speechgpt,Connecting_Encoder_LLM,Qwen3-ASR}. However, standard LLMs are
designed to generate text autoregressively and do not have a natural mechanism
to control when a token should be emitted while speech is continuously arriving.

Delayed Streams Modeling (DSM) addresses this problem by using the same
decoder-only LLM to process time-aligned acoustic and text streams
\cite{kyutai_DSM,Voxtral_realtime}. In addition to normal text tokens, DSM adds
two special tokens to the LLM vocabulary. The padding token \texttt{<p>} means
that the model should wait without producing text, while the word-start token
\texttt{<w>} marks the beginning of a word. These special tokens are predicted
together with normal text tokens using the same softmax.

This simple design allows an LLM to perform streaming ASR, but it also places
ASR-specific timing tokens and normal text tokens in the same output
distribution. In a time-aligned sequence, many steps have no text output and
therefore use \texttt{<p>} as the target. As a result, \texttt{<p>} appears much
more often than any individual text token. Still, DSM computes the projection
and softmax over the full vocabulary even when the model only needs to wait. In
addition, \texttt{<p>} and \texttt{<w>} are included in the softmax
normalization, so these ASR-specific tokens compete with the distribution over
the original LLM vocabulary learned during pretraining. Streaming ASR therefore
involves two distinct decisions: whether to emit a token and, if so, which token
to emit. DSM couples these decisions in a single softmax, motivating us to model
timing and lexical prediction separately.

We first study whether \texttt{<w>} is necessary for speech recognition. Our
experiments show that removing \texttt{<w>} still gives competitive recognition
performance. This result suggests that the ASR-specific output space of DSM can
be simplified to only one special token, \texttt{<p>}. The remaining
\texttt{<p>} can be viewed as a timing symbol rather than a normal text token,
similar to the blank symbol in CTC and RNN-T. Based on this view, we propose
Factorized DSM (F-DSM), inspired by the Hybrid Autoregressive Transducer (HAT)
\cite{HAT}. F-DSM separates the waiting probability for \texttt{<p>} from the
distribution over the original LLM vocabulary. With this factorization, the text
prediction head keeps the same output space as the pretrained LLM and does not
include any ASR-specific token. In addition, when the model decides to wait, it
does not need to compute the large-vocabulary projection and softmax. F-DSM
therefore separates ASR-specific timing prediction from the original text-token
prediction of the LLM while keeping the streaming framework of DSM.

We evaluate F-DSM on Japanese and English using the Corpus of Spontaneous
Japanese (CSJ) \cite{CSJ} and LibriSpeech \cite{LibriSpeech}. F-DSM achieves
better recognition performance than the original DSM. During training, it
greatly reduces GPU memory use while keeping similar training throughput,
resulting in more memory-efficient training. During inference, skipping the
large-vocabulary softmax on waiting steps provides a small speed improvement. We
also evaluate text-only language modeling. Compared with the original DSM, F-DSM
reduces the degradation in perplexity (PPL) of the pretrained LLM, showing that
separating ASR-specific tokens from the original LLM vocabulary better preserves
its language modeling behavior.

These results show that treating \texttt{<p>} separately from normal text tokens
is useful not only for reducing computation but also for preserving the original
language modeling ability of the LLM. The factorized design requires only a
small change to DSM while keeping its streaming framework unchanged. This makes
F-DSM a simple and effective approach for LLM-based streaming ASR.

\section{\revisiontext{Preliminaries}}
\label{sec:preliminaries}

\subsection{\revisiontext{Delayed Streams Modeling}}
\label{sec:prelim-dsm}

DSM \cite{kyutai_DSM} is a method for streaming speech recognition with a
decoder-only LLM. As shown on the left side of Fig.~\ref{fig:method-overview},
DSM aligns acoustic and text streams in time and shifts the text stream by a
fixed delay. This allows the model to predict the next token sequentially using
acoustic information and previous text.

In addition to normal text tokens, DSM adds two special tokens to the LLM
vocabulary. The padding token \texttt{<p>} represents waiting without producing
text, while the word-start token \texttt{<w>} marks the beginning of a word. Let
$\mathcal{V}$ denote the original LLM vocabulary. The output vocabulary of DSM
is defined as
\begin{equation}
    \mathcal{V}_{\mathrm{DSM}}
    =
    \mathcal{V}
    \cup
    \{\texttt{<p>}, \texttt{<w>}\}.
\end{equation}
\begingroup
Let $t$ index aligned time frames, $y_t$ be the output token, and
$\mathbf{h}_t\in\mathbb{R}^{d}$ be the decoder hidden state. DSM predicts each
token $v\in\mathcal{V}_{\mathrm{DSM}}$ as
\begin{equation}
    P_{\mathrm{DSM}}(y_t=v\mid\mathbf{h}_t)
    =\mathrm{Softmax}(\mathbf{W}_{\mathrm{DSM}}\mathbf{h}_t)_v,
    \label{eq:dsm-output}
\end{equation}
where $\mathbf{W}_{\mathrm{DSM}}\in\mathbb{R}^{(|\mathcal{V}|+2)\times d}$. This
full-vocabulary projection is computed at every frame, including waiting frames.
\par\endgroup

{The word-start token \texttt{<w>} was introduced to explicitly mark the
transition from padding to text and support timestamp extraction during
inference~\cite{kyutai_DSM,Voxtral_realtime}. However, its necessity has not
been established through an ablation that removes it. ASR systems can provide
word-level timestamps without a dedicated word-start token~\cite{WhisperX},
suggesting that
\texttt{<w>} may also be unnecessary in DSM.
This motivates our ablation of \texttt{<w>} for recognition accuracy.}

\subsection{\revisiontext{Hybrid Autoregressive Transducer}}
\label{sec:prelim-hat}

{HAT~\cite{HAT} factorizes the output distribution of RNN-T into a blank
probability and a conditional distribution over non-blank labels
\cite{F-Neural_Transducer,F-Transducer}. This separates the decision of whether
to emit a label from the decision of which label to emit.}

\begingroup
At acoustic frame $t$ after $u$ emitted labels, let $\mathbf{h}_{t,u}$ be HAT's
joint hidden state, $b_{t,u}$ its blank probability, and $q_{t,u}(v)$ its
conditional distribution over non-blank labels $v\in\mathcal{V}$. Then
\begin{equation}
    P_{\mathrm{HAT}}(v\mid\mathbf{h}_{t,u})=
    \begin{cases}
        b_{t,u}, & v=\texttt{<blank>},\\[2mm]
        (1-b_{t,u})q_{t,u}(v), & v\in\mathcal{V}.
    \end{cases}
    \label{eq:hat-output}
\end{equation}
\endgroup
{This factorization motivates treating the waiting token \texttt{<p>} separately
from text tokens in DSM.}

\section{Factorized Delayed Streams Modeling}
\label{sec:method}

\subsection{Overview}

{Building on the DSM framework and the HAT factorization described in
Sec.~\ref{sec:preliminaries}, we propose F-DSM, which removes \texttt{<w>} and
separates the waiting probability for \texttt{<p>} from the distribution over
the original LLM vocabulary.}

{Following HAT and RNN-T, we feed back only emitted text tokens to the LLM
decoder for autoregressive prediction, excluding the waiting token
\texttt{<p>}.}

\subsection{Factorized Output Distribution}

{Let $\mathbf{h}_t\in\mathbb{R}^{d}$ denote the LLM decoder hidden state at time
frame $t$, as in DSM.} F-DSM first computes the waiting probability $b_t$ using
the padding prediction head as
\begin{equation}
    b_t
    =
    \sigma
    \left(
    \mathbf{w}_{p}^{\top}\mathbf{h}_t + c_p
    \right),
\end{equation}
{where $\sigma$ is the sigmoid function, $\mathbf{w}_p\in\mathbb{R}^{d}$ is a
learned weight vector, and $c_p\in\mathbb{R}$ is a bias. Thus, the padding head
is a scalar linear layer followed by a sigmoid.}

For normal text tokens, the original LLM head predicts a distribution over the
original LLM vocabulary $\mathcal{V}$ as
\begin{equation}
    P_{\mathrm{LLM}}
    (v \mid \mathbf{h}_t)
    =
    \mathrm{Softmax}
    \left(
    \mathbf{W}_{\mathrm{LLM}}\mathbf{h}_t
    \right)_v,
    \quad
    v \in \mathcal{V}.
    \label{eq:llm-head}
\end{equation}
{Here, $\mathbf{W}_{\mathrm{LLM}}\in\mathbb{R}^{|\mathcal{V}|\times d}$ is the
pretrained LLM output matrix. Its projection costs $O(|\mathcal{V}|d)$
operations per frame, versus $O(d)$ for the padding head. Unlike DSM in
Eq.~\ref{eq:dsm-output}, F-DSM can skip this expensive projection at waiting
positions.}

{In contrast to the joint softmax in Eq.~\ref{eq:dsm-output}, F-DSM defines its
output probability as}
\begin{equation}
    P_{\mathrm{F\mathchar`-DSM}}
    (y_t \mid \mathbf{h}_t)
    =
    \begin{cases}
        b_t,
        & y_t = \texttt{<p>}, \\[2mm]
        (1-b_t)
        P_{\mathrm{LLM}}(y_t \mid \mathbf{h}_t),
        & y_t \in \mathcal{V}.
    \end{cases}
    \label{eq:fdsm-output}
\end{equation}
{This follows the HAT factorization in Eq.~\ref{eq:hat-output}, with
\texttt{<p>} serving as the waiting symbol.} In this formulation, \texttt{<p>}
is not included in the original LLM vocabulary but is modeled separately as the
waiting probability. Therefore, the LLM head keeps the same output space as in
pretraining.

\subsection{Training and Streaming Inference}

F-DSM is trained using cross-entropy loss with the factorized output probability
defined above. The loss at step $t$ is written as
\begin{equation}
    \mathcal{L}_t
    =
    \begin{cases}
        -\log b_t,
        & y_t = \texttt{<p>}, \\[2mm]
        -\log(1-b_t)
        -\log
        P_{\mathrm{LLM}}(y_t \mid \mathbf{h}_t),
        & y_t \in \mathcal{V}.
    \end{cases}
\end{equation}
For \texttt{<p>} targets, the vocabulary projection and softmax in
Eq.~\ref{eq:llm-head} can be skipped during training, reducing memory use for
large-vocabulary logits.

During streaming inference, the padding prediction head first computes $b_t$.
Similar to blank-thresholding approaches for HAT-based
ASR~\cite{F-Transducer,moriya24_interspeech}, F-DSM uses the waiting probability
to avoid expensive non-waiting computation. Specifically, under greedy decoding,
F-DSM skips the LLM vocabulary projection and softmax when the waiting
probability exceeds 0.5:
\begin{equation}
    b_t > 0.5 \quad \Longrightarrow \quad {\text{skip Eq.~\ref{eq:llm-head}}}.
    \label{eq:softmax-skipping}
\end{equation}
At these steps, the model outputs \texttt{<p>} and moves to the next speech
step.

\section{Experiments}
\label{sec:experiments}

\subsection{Experimental Setup}

We use CSJ for Japanese experiments and LibriSpeech for English experiments. The
base LLM is Qwen3.5-0.8B \cite{qwen3.5}, and the acoustic encoder is the
Fast-Conformer \cite{FastConformer} encoder from Nemotron-3.5-ASR-Streaming-0.6B
\cite{cache_aware_streaming, nvidia_nemotron35_asr}. The encoder operates at a
frame rate of 12.5~Hz. A randomly initialized two-layer MLP is used as an audio
projection adapter between the acoustic encoder and the LLM.

The speech and text in both datasets are aligned using Qwen3-Aligner
\cite{Qwen3-ASR}, and the alignments are represented at the 12.5-Hz frame rate
of the acoustic encoder. During training, the target delay is uniformly sampled
from 160, 320, 480, 640, and 800~ms. Following Voxtral Realtime, we use Ada
RMSNorm to condition the model on the target delay \cite{Voxtral_realtime}.

We compare DSM w/ \texttt{<w>}, DSM w/o \texttt{<w>}, and F-DSM. The second
removes only \texttt{<w>} from DSM, while F-DSM further factorizes the
prediction of \texttt{<p>} as described in Sec.~\ref{sec:method}.

For the two DSM baselines, we apply z-loss regularization, following the setting
reported to be effective in Voxtral Realtime \cite{Voxtral_realtime}. F-DSM does
not use z-loss regularization because \texttt{<p>} is not included in the LLM
vocabulary.

Training is performed on {four} NVIDIA A100 GPUs with a global batch size of
128. Each model is trained for a total of 100,000 steps using a three-stage
training schedule. For the first 5,000 steps, only the audio projection adapter
and the newly added prediction parameters are updated. For the next 5,000 steps,
the acoustic encoder is additionally updated. For the remaining 90,000 steps,
the acoustic encoder, adapter, prediction parameters, and LoRA \cite{LoRA}
parameters are updated. The base LLM parameters remain frozen, and the LLM is
adapted using LoRA.

We use AdamW \cite{AdamW} with zero weight decay. For all models, we use the
learning-rate setting that {gives} the best performance for DSM w/ \texttt{<w>}.
The learning rates are \(3\times10^{-4}\) for the language-model adapters,
\(10^{-5}\) for the acoustic encoder, \(10^{-4}\) for the audio projection and
delay-conditioning modules, and
\(5\times10^{-5}\) for the streaming-token parameters.

Recognition performance is evaluated using greedy decoding, with LLM-head
skipping according to Eq.~\ref{eq:softmax-skipping} for F-DSM. We use character
error rate (CER) for CSJ and word error rate (WER) for LibriSpeech.

\subsection{Speech Recognition Accuracy}

\begin{table}[t]
    \centering
    \caption{ASR error with greedy decoding, 480~ms target delay,
    and no left padding (lower is better).}
    \label{tab:asr}
    \small
    \setlength{\tabcolsep}{1.5pt}
    \begin{tabular*}{\columnwidth}{@{\extracolsep{\fill}}lrrrrrrr@{}}
        \toprule
        & \multicolumn{4}{c}{CSJ (CER, \%)}
        & \multicolumn{3}{c}{LibriSpeech (WER, \%)} \\
        \cmidrule(lr){2-5}\cmidrule(lr){6-8}
        Model & eval1 & eval2 & eval3 & Ave.
        & test-clean & test-other & Ave. \\
        \midrule
        DSM w/ \texttt{<w>}
        & 7.91 & 6.80 & 7.08 & 7.27
        & 3.86 & 9.17 & 6.52 \\
        DSM w/o \texttt{<w>}
        & 7.78 & 6.46 & 5.54 & 6.60
        & 3.83 & 9.38 & 6.61 \\
        F-DSM
        & \textbf{7.54} & \textbf{6.29} & \textbf{5.50}
        & \textbf{6.45}
        & \textbf{3.66} & \textbf{8.68} & \textbf{6.17} \\
        \bottomrule
    \end{tabular*}
\end{table}

Table~\ref{tab:asr} compares the three models at a target delay of 480~ms.
Removing \texttt{<w>} improves CER on CSJ while changing WER only slightly on
LibriSpeech, indicating that explicit word-start prediction is not required for
competitive recognition in our setting. F-DSM further improves the average CER
from 6.60\% to 6.45\% on CSJ and the average WER from 6.61\% to 6.17\% on
LibriSpeech.

\begin{figure}[t]
    \centering
    \includegraphics[width=0.88\columnwidth,trim=0 70 0 110,clip]{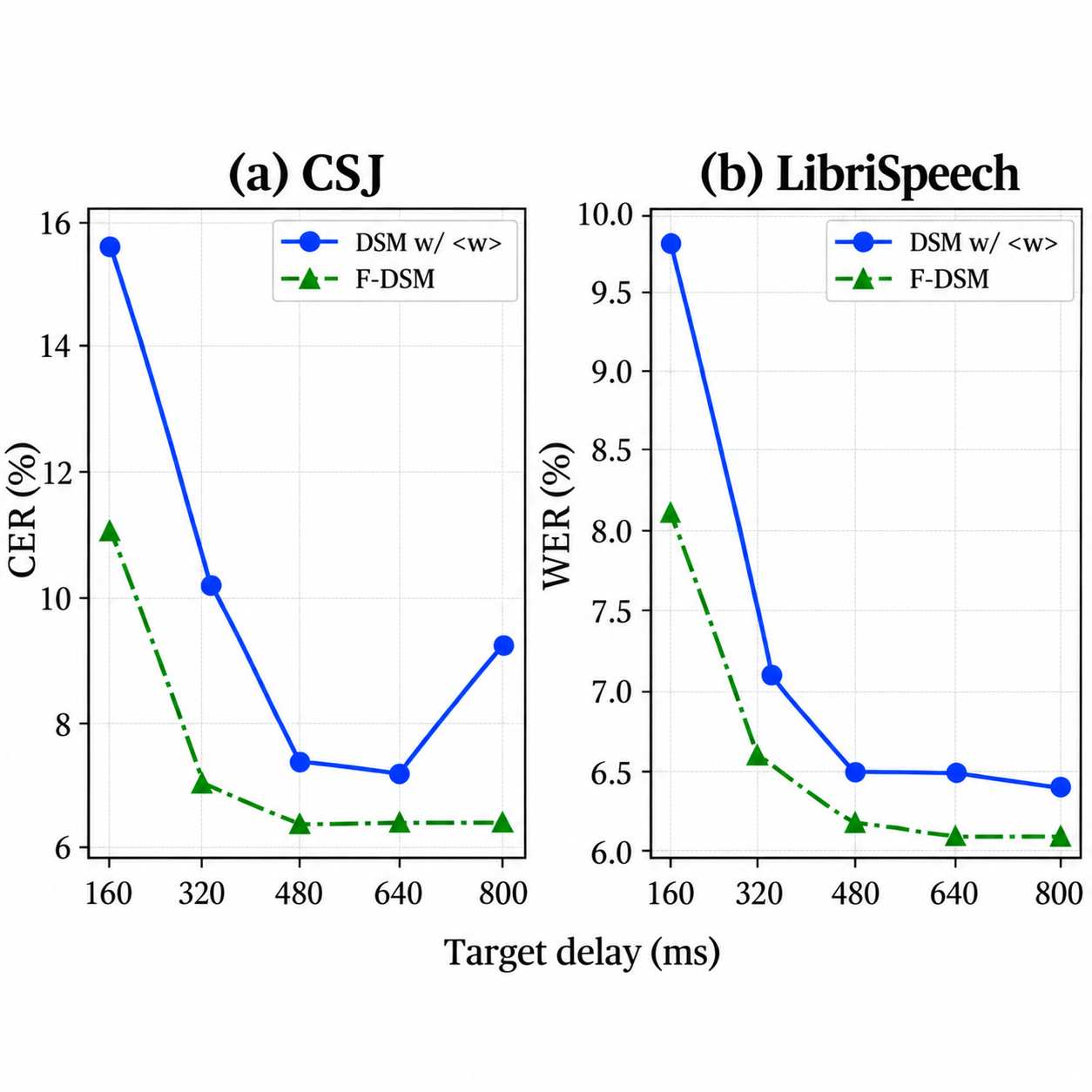}
    \caption{Error versus target delay: (a) CSJ CER and
    (b) LibriSpeech WER (lower is better).}
    \label{fig:cer-wer-delay}
\end{figure}

{Fig.~\ref{fig:cer-wer-delay} compares F-DSM with DSM w/ \texttt{<w>} at target
delays of 160--800~ms. F-DSM achieves lower error rates at every evaluated delay
on both datasets, with the largest gap at 160~ms. For F-DSM, error rates
decrease as the delay increases to 480~ms, with little further change at longer
delays. On CSJ, DSM's CER rises at 800~ms, whereas F-DSM's CER remains stable.
These results show that F-DSM improves recognition across the evaluated delay
settings, including the shortest target delay.}

\subsection{Training Efficiency}

\begin{table}[t]
    \centering
    \caption{Training efficiency at batch sizes 16 / 32 / 64.
    OOM means out of memory. --- means VRAM unavailable due to OOM.}
    \label{tab:efficiency}
    \small
    \setlength{\tabcolsep}{2pt}
    \renewcommand{\arraystretch}{1.08}
    \begin{tabular*}{\columnwidth}{@{\extracolsep{\fill}}lcc@{}}
        \toprule
        Model & Time (s/step) $\downarrow$ & Ave. VRAM (GiB) $\downarrow$ \\
        \midrule
        \multicolumn{3}{@{}l}{\textit{CSJ (mean utterance: 4.63 s)}} \\
        \addlinespace[2pt]
        DSM w/ \texttt{<w>}
        & \textbf{0.886} / \textbf{1.309} / OOM
        & 35.14 / 61.30 / --- \\
        DSM w/o \texttt{<w>}
        & 1.026 / 1.427 / OOM
        & 35.09 / 60.95 / --- \\
        F-DSM
        & 1.043 / 1.375 / 2.478
        & \textbf{14.71} / \textbf{18.49} / \textbf{25.58} \\
        \midrule
        \multicolumn{3}{@{}l}{\textit{LibriSpeech (mean utterance: 12.30 s)}} \\
        \addlinespace[2pt]
        DSM w/ \texttt{<w>}
        & 1.178 / 1.771 / OOM
        & 38.34 / 65.82 / --- \\
        DSM w/o \texttt{<w>}
        & 1.133 / 1.771 / OOM
        & 37.49 / 64.66 / --- \\
        F-DSM
        & \textbf{1.100} / \textbf{1.714} / 3.008
        & \textbf{18.04} / \textbf{24.51} / \textbf{38.69} \\
        \bottomrule
    \end{tabular*}
\end{table}

Table~\ref{tab:efficiency} compares training time and GPU memory use. {As
discussed in Sec.~\ref{sec:method},} in F-DSM, the large LLM head does not need
to be evaluated at positions where the target is \texttt{<p>}, reducing the
memory required for large-vocabulary logits.

At batch size 32, DSM w/o \texttt{<w>} uses 60.95~GiB on CSJ, while F-DSM uses
only 18.49~GiB. On LibriSpeech, GPU memory use is reduced from 64.66~GiB to
24.51~GiB. At the same time, the training time per step remains similar: 1.427~s
versus 1.375~s on CSJ and 1.771~s versus 1.714~s on LibriSpeech.

{On CSJ, whose mean utterance length is 4.63~s, DSM w/ \texttt{<w>} is faster,
likely because it avoids the overhead of selectively skipping vocabulary
computation. For the longer utterances in LibriSpeech (12.30~s on average), this
overhead is less significant relative to the total computation, and F-DSM is
slightly faster at matched batch sizes. We expect longer utterances to further
reduce the relative impact of this overhead and make the memory savings more
beneficial.}

Both DSM baselines run out of memory at batch size 64, whereas F-DSM can be
trained at this batch size on both datasets. {This allows larger batches to be
processed directly, reducing the need for gradient accumulation and enabling
more efficient training.}

\subsection{Inference Efficiency}

\begin{table}[t]
    \centering
    \caption{Streaming inference efficiency (CSJ / LibriSpeech).
    Skip rate: decoding steps skipping vocabulary projection and
    softmax (\%). RTF: processing time/audio duration.
    --: not applicable.}
    \label{tab:inference-efficiency}
    \small
    \setlength{\tabcolsep}{6pt}
    \begin{tabular}{@{}lcc@{}}
        \toprule
        Model & \shortstack{Softmax skip\\rate (\%)} & RTF $\downarrow$ \\
        \midrule
        DSM w/ \texttt{<w>}  & -- / -- & 0.827 / 0.780 \\
        DSM w/o \texttt{<w>} & -- / -- & 0.824 / 0.764 \\
        F-DSM                & 74.70 / 66.89 & \textbf{0.811} / \textbf{0.759} \\
        \bottomrule
    \end{tabular}
\end{table}

{As described in Sec.~\ref{sec:method},} during streaming inference, F-DSM first
uses the padding prediction head to determine whether the model should output
\texttt{<p>}. When \texttt{<p>} is selected, the projection and softmax over the
large LLM vocabulary can be skipped.

In our experiments, F-DSM skips the large-vocabulary softmax on 74.70\% of
decoding steps on CSJ and 66.89\% on LibriSpeech, as shown in
Table~\ref{tab:inference-efficiency}. Thus, F-DSM avoids vocabulary computation
on most decoding steps. F-DSM achieves the lowest RTF on both datasets (0.811 on
CSJ and 0.759 on LibriSpeech), indicating a modest improvement in end-to-end
inference speed. The gain is limited because the acoustic encoder and LLM
decoder still need to be computed.

\subsection{\revisiontext{Analysis of LLM Forgetting}}

\begin{table}[t]
    \centering
    \caption{Text-only language modeling after ASR adaptation.
    {Renorm. PPL excludes special tokens and renormalizes
    probabilities over the original vocabulary.}
    Special-token probability: summed raw-head probabilities of
    \texttt{<p>} or \texttt{<w>}.}
    \label{tab:ppl}
    \small
    \setlength{\tabcolsep}{2pt}
    \begin{tabular*}{\columnwidth}{@{\extracolsep{\fill}}lrrr@{}}
        \toprule
        Model & Raw PPL $\downarrow$
        & {\shortstack{Renorm.\\PPL $\downarrow$}}
        & \shortstack{Special-token\\probability (\%)} \\
        \midrule
        \multicolumn{4}{@{}l}{\textit{CSJ}} \\
        Base Qwen & {$4.401\times10^{2}$} & --- & --- \\
        DSM w/ \texttt{<w>} & {$1.858\times10^{6}$}
        & {$3.293\times10^{5}$} & {51.77} \\
        DSM w/o \texttt{<w>} & {$3.113\times10^{6}$}
        & {$2.407\times10^{5}$} & {77.97} \\
        F-DSM & {\boldmath $2.388\times10^{5}$}
        & {\boldmath $2.388\times10^{5}$} & \textbf{0.002183} \\
        \midrule
        \multicolumn{4}{@{}l}{\textit{LibriSpeech}} \\
        Base Qwen & {$5.945\times10^{1}$} & --- & --- \\
        DSM w/ \texttt{<w>} & {$2.290\times10^{5}$}
        & {$5.003\times10^{4}$} & {55.94} \\
        DSM w/o \texttt{<w>} & {$4.831\times10^{5}$}
        & {$4.692\times10^{4}$} & {76.95} \\
        F-DSM & {\boldmath $4.296\times10^{4}$}
        & {\boldmath $4.296\times10^{4}$} & \textbf{0.001579} \\
        \bottomrule
    \end{tabular*}
\end{table}

Fine-tuning LLMs on downstream tasks can lead to catastrophic forgetting of
previously learned capabilities~\cite{Catastrophic_Forgetting_LLM}. We therefore
evaluate text-only PPL to investigate how ASR adaptation affects the original
language modeling capability of the LLM. The results are shown in
Table~\ref{tab:ppl}.

{F-DSM achieves the lowest raw PPL among the adapted models on both datasets,
substantially mitigating the degradation observed with DSM.} The DSM baselines
also assign a large amount of probability mass to ASR-specific tokens during
text-only evaluation. On CSJ, the probability mass assigned to special tokens is
{51.77\%} for DSM w/ \texttt{<w>} and {77.97\%} for DSM w/o \texttt{<w>},
whereas it is only 0.002183\% for F-DSM. A similar trend is observed on
LibriSpeech.

{Even after renormalizing the probability distribution over text tokens only,
F-DSM achieves the lowest PPL. However, its advantage over both DSM baselines
shrinks substantially, and all adapted models still have much higher PPL than
the original LLM. These results suggest that gradients from \texttt{<p>}
prediction may not be completely isolated from the LLM, even in F-DSM. Jointly
providing audio and text embeddings to the LLM may also degrade its language
modeling capability.}

\section{Conclusion}
\label{sec:conclusion}

We proposed F-DSM, which separates the waiting probability for \texttt{<p>} from
text-token prediction over the original LLM vocabulary. Experiments on CSJ and
LibriSpeech showed that the word-start token
\texttt{<w>} is not required for competitive recognition accuracy and
that F-DSM outperforms the original DSM. F-DSM also greatly reduces GPU memory
use while maintaining similar training throughput, and skips the
large-vocabulary softmax on 74.70\% of decoding steps on CSJ and 66.89\% on
LibriSpeech with modest gains in end-to-end inference speed. In addition, F-DSM
reduces the degradation in text-only PPL observed with DSM, although a gap from
the base LLM remains. Separating waiting from text prediction thus improves
LLM-based streaming ASR. Future work will explore mechanisms to skip LLM
inference itself, aiming to improve performance by further mitigating
catastrophic forgetting and to reduce computational cost.

\newpage
\begingroup
\apptocmd{\thebibliography}{\setlength{\itemsep}{1pt}\setlength{\parsep}{0pt}}{}{}
\bibliographystyle{IEEEbib}
\nocite{*}
\bibliography{refs}
\endgroup

\end{document}